\documentclass[aps,preprint]{revtex4}%
\usepackage{amsfonts}
\usepackage{amsmath}
\usepackage{amssymb}
\usepackage{graphicx}%
\providecommand{\U}[1]{\protect\rule{.1in}{.1in}}

\begin{document}
\author{}
\title{Spectroscopy of a weakly isolated horizon}
\author{Ge-Rui Chen}\email{chengerui@emails.bjut.edu.cn}
\author{Yong-Chang Huang}
\affiliation{Institute of Theoretical Physics, Beijing University of
Technology, Beijing, 100124, China}

\begin{abstract}
The spectroscopy of a weakly isolated horizon (WIH) has been investigated. We obtain an equally spaced entropy
spectrum with its quantum equal to the one given by
Bekenstein \cite{jdb}. We demonstrate that the quantization of entropy and area
is a generic property of horizon, and the results exit in a wide class of
spacetimes admitting weakly isolated horizons. Our results also indicate that the entropy quantum of the black hole horizon is closely related to Hawking temperature.
\end{abstract}

\maketitle

\section{\textbf{Introduction}}

Since the first exact solution of Einstein equation was found out,
studying black holes' properties has become an important part of
gravitational physics. Properties of black holes, for example, laws
of black hole mechanics, Hawking radiation and black hole
spectroscopy, caused deep, unsuspected connections among classical
general relativity, quantum physics and statistical mechanics. However, the traditional definition of a
black hole \cite{rmw}, is too global and idealized: it requires
knowledge of the entire future of the space-time, so it is often
cumbersome to use for the requirements of practical research
\cite{aa7}. In recent years, a new, quasi-local framework was
introduced by Ashtekar and his collaborators to analyze different
facets of black holes in a unified way \cite{aa2,aa5,aa7}. Compared
with the event horizon, this framework doesn't need the knowledge of
overall space-time, and only involves quasi-local conditions, so it
accords with the practical physical process. In this framework,
black holes in equilibrium (no matter and energy flow across the
horizon) are described by weakly isolated horizons (WIH).

In 1970s, Bekenstein proved that the quantum of the black hole
horizon is given as $(\Delta A)_{min}=8\pi l_p^2$ \cite{jdb}. From
then on, there has been much attention paid to the quantization of
black hole entropy spectrum and area spectrum
\cite{sh,gk,vfm,mm,ecv,ajmm,rbj1,rbj2,rbj3,kr}, and many methods rely on the quasi-normal frequency which requires the
knowledge of the global geometry of the space-time, not just the
geometry near a horizon. Recently, Majhi and Vagenas \cite{bec}
proposed a new approach to derive the entropy spectrum and the
horizon area quantum utilizing solely the periodicity of imaginary time and the Bohr-Sommerfekd quantization rule, and there was no use
at all of the quasinormal frequencies to obtain the result. Later on,
there were many works using this method to study the entropy
spectrum and area spectrum of a wide variety of spacetimes \cite{xxz}.  We use a similar method that does not need the periodicity of imaginary time, and our method could applies to more general spacetimes. In this paper, we quantize the  spectroscopy of a weakly isolated horizon which is a locally defined black hole and already contains all the stationary black holes. Our method also shows that there exists close relationship between the entropy quantum and Hawking temperature, which is an interesting thing.

This paper is organized as follows. In section $2$, we briefly
review the definition of the WIH and the geometry
near it. In
section $3$, we quantize a weakly isolated horizon. Finally,
the conclusions are given in section $4$.

\section{Geometry of a weakly isolated horizon}
In this section we will briefly review some geometric properties of
WIH \cite{aa2,aa5,aa7}. As in Refs. \cite{bk}, it is very
convenient to introduce the Bondi-like coordinates
$(u,r,\theta,\varphi)$, which are well defined on the horizon, and
choose a set of null tetrad, which satisfy Bondi gauge, to study the
behavior in the neighborhood of WIH. The null tetrad can be
expressed as
\begin{eqnarray}
l^{a}&=&\frac{\partial}{\partial u}+U\frac{\partial}{\partial
r}+X\frac{\partial}{\partial
\varsigma}+\overline{X}\frac{\partial}{\partial\overline{\varsigma}}\nonumber\\
n^{a}&=&-\frac{\partial}{\partial r}\nonumber\\
m^{a}&=&\omega \frac{\partial}{\partial
r}+\xi_{3}\frac{\partial}{\partial
\varsigma}+\xi_{4}\frac{\partial}{\partial \overline{\varsigma}}\nonumber\\
\overline{m}^{a}&=&\overline{\omega}\frac{\partial}{\partial
r}+\overline{\xi}_{3}\frac{\partial}{\partial
\overline{\varsigma}}+\overline{\xi}_{4}\frac{\partial}{\partial
\varsigma},\label{biaojia}
\end{eqnarray}
where $U\ \widehat{=}\ X\ \widehat{=}\ \omega\ \widehat{=}\ 0 $ on
the horizon $H$ (following the notation in Ref. \cite{aa7},
equalities restricted to $H$ will be denoted by `` $\widehat{=}$ "),
and $\varsigma =e^{i\phi}cot\frac{\theta}{2}$. Note that $n^{a}$ and
$l^{a}$ are future directed. We take the spacetime metric $g_{ab}$
to have a signature $(-,+,+,+)$, so the metric can be expressed as
\begin{eqnarray}
g_{ab}=m_{a}\overline{m}_{b}+\overline{m}_{a}m_{b}-n_{a}l_{b}-l_{a}n_{b}.\label{dg}
\end{eqnarray}

The definition of WIH \cite{aa2,aa5,aa7} implies that there is a one
form $\omega_{a}$ on $H$ which satisfy the following relationship,
$\pounds_{l}\omega^{a}\widehat{=}0$ and
$D_{a}l^{b}\widehat{=}\omega_{a}l^{b}$, where $D_{a}$ is the induced
covariant derivative on $H$. In terms of the Newman-Penrose
formalism, $\omega_{a}$ can be explicitly expressed as
\begin{eqnarray}
\omega_{a}=-(\varepsilon+\overline{\varepsilon})n_{a}+
(\alpha+\overline{\beta})\overline{m}_{a}+(\overline{\alpha}+\beta)m_{a}=
-(\varepsilon+\overline{\varepsilon})n_{a}+\pi\overline{m}_{a}+\overline{\pi}m_{a},\label{bj6}
\end{eqnarray}
which means $(\varepsilon+\overline{\varepsilon})$ is constant on
$H$ from $\pounds_{l}\omega^{a}\widehat{=}0$. The commutators of the
null tetrad $[l^{a}, n^{a}]$ and $[m^{a}, n^{a}]$ tell us that
\begin{eqnarray}
\frac{\partial U}{\partial
r}&=&(\varepsilon+\overline{\varepsilon})+\overline{\pi}\overline{\omega}+\pi\omega
,\ \frac{\partial X}{\partial
r}=\overline{\pi}\overline{\xi}_{4}+\pi\xi_{3},\ \frac{\partial
\omega}{\partial
r}=\overline{\pi}+\overline{\lambda}\overline{\omega}+\mu\omega,\nonumber\\
\frac{\partial \xi_{3}}{\partial
r}&=&\overline{\lambda}\overline{\xi}_{4}+\mu\xi_{3},\
\frac{\partial\xi_{4}}{\partial
r}=\overline{\lambda}\overline{\xi}_{3}+\mu\xi_{4},
\end{eqnarray}
which means $\frac{\partial U}{\partial r}\ \widehat{=}\
(\varepsilon+\overline{\varepsilon})$. Then the behavior of
functions $U$, $X$ and $\omega$ near $H$ is
\begin{eqnarray}
U&=&(\varepsilon+\overline{\varepsilon})r+O(r^2)\nonumber,\\
X&=&O(r), \omega=O(r).\label{dg2}
\end{eqnarray}
Using Eq. (\ref{biaojia}), Eqs. (\ref{dg}) and Eq. (\ref{dg2}), the
out-going null geodesic can be calculated as
\begin{eqnarray}
0=2du^2\frac{\overline{\xi}_{3}X-\overline{\xi}_{4}\overline{X}}{|\xi_{4}|^{2}-|\xi_{3}|^{2}}\times\frac{\xi_{3}\overline{X}-\xi_{4}X}{|\xi_{4}|^{2}-|\xi_{3}|^{2}}
-2du^2(U-\frac{\xi_{4}\overline{\omega}-
\overline{\xi}_{3}\omega}{|\xi_{4}|^{2}-|\xi_{3}|^{2}}X-\frac{\overline{\xi}_{4}\omega-\xi_{3}\overline{\omega}}{|\xi_{4}|^{2}-|\xi_{3}|^{2}}\overline{X})+2dudr,
\end{eqnarray}
which leads to
\begin{eqnarray}
\frac{dr}{du}=U\label{rn1}.
\end{eqnarray}

Based on Ref. \cite{aa7}, not any choice of time direction can give
a Hamiltonian evolution, and only some suitably chosen time
direction can lead to a well-defined horizon mass. In Ref.
\cite{aa7}, A. Ashtekar and B. Krishnan gave a canonical way to
choose the time direction $t^a$ for a WIH, and the restriction of
$t^a$ to $H$ should be a linear combination of a null normal $l^a$
and the axisymmetric vector $\psi^a$,
\begin{eqnarray}
t^a\ \widehat{=}\ B_tl^a-\Omega_{t}\psi^a\label{rn2},
\end{eqnarray}
where $B_t$ and $\Omega_{t}$ are constant on the horizon. Compared
with the Schwarzschild case, the parameter of $t^a$ takes the place
of the Killing time. Using Eq. (\ref{rn1}) and Eq. (\ref{rn2}), we
get the time derivative of $r$ along the outgoing geodesic,
\begin{eqnarray}
\dot{r}=\frac{du}{dt}\frac{dr}{du}=(B_t+O(r))U=B_t(\varepsilon+\overline{\varepsilon})r+O(r^2)\label{bj1}.
\end{eqnarray}

With the canonical time direction $t^a$, A. Ashtekar and B. Krishnan
\cite{aa7} established the zeroth and the first law of WIH. By
definition, the surface gravity of $H$ is
$\kappa_{t}:=B_tl^a\omega_a=B_t(\varepsilon+\overline{\varepsilon})$.
Because $B_t(\varepsilon+\overline{\varepsilon})$ is constant on
$H$, the zeroth law of black hole mechanics is valid for WIH. The
first law is expressed as
\begin{eqnarray}
\delta M_{H}^{(t)}=\frac{\kappa_{t}}{8\pi}\delta
a_{H}+\Omega_{t}\delta J_{H}\label{jh},
\end{eqnarray}
where $M_{H}^{(t)} $ is the horizon mass, $ a_H $ is the area of the
cross section of WIH, $\Omega_{t}$ is the angular velocity of the
horizon and $J_H=-\frac{1}{8\pi}\oint_S(\omega_a\psi^a)dS$ is the
angular momentum. The first law of WIH is the generalization of the
first law of stationary black holes. Because Refs. \cite{xnw1,xnw2}
studied the Hawking radiation of a WIH, so the first law of WIH mechanics
is upgraded to the first law of WIH thermodynamics which can be expressed as
\begin{eqnarray}
\delta M_{H}^{(t)}=T_H\delta S_H+\Omega_{t}\delta J_{H},\label{q1}
\end{eqnarray}
where $S_H=\frac{a_H}{4}$ is the entropy of WIH, and $T_H=\frac{\kappa_{t}}{2\pi}$ is the Hawking temperature.

\section{quantization of a weakly isolated horizon }
In this section, we quantize a weakly isolated
horizon. We consider a tunneling process cross the WIH, and investigate an action of the
form
\begin{equation}
I=\int p_i
dq_i=\int\int dp_{i}dq_{i}=\int\int\frac{d\varepsilon}{\dot{q_i}}dq_i,\label{b1}
\end{equation}
where $p_i$ is the conjugate momentum of the coordinate $q_i$ with
$i=0,1$ for which  $q_0=\tau$ and $q_1=r$. Note that we use
the Euclidean time $q_0=\tau$ and the Einstein summation convention.
To get the last equation, we have used Hamiltion' s equation
$\dot{q_i}=\frac{d\varepsilon}{dp_i}$, where the Hamiltonian $\varepsilon$ is the energy of the emitted particle. Write Eq. (\ref{b1}) explicitly,
\begin{equation}
I=\int p_i
dq_i=\int\int d\varepsilon d\tau+\int\int\frac{d\varepsilon}{\dot{r}}dr\label{rn3}.
\end{equation}
We shall obtain the quantity $\dot{r}$ that appears in Eq. (\ref{rn3}).
Let us consider the radial null path, and our analysis will concentrates on the outgoing path, since
these is the one related to the quantum mechanically nontrivial
features \cite{mk1}. Since in Eq. (\ref{rn3}) $\tau$ is the Euclidean time, we substitute the transformation $t\rightarrow i\tau$ into Eq.
(\ref{bj1}) and get the outgoing radial null path. This leads to
\begin{eqnarray}
\dot{r}\equiv
\frac{dr}{d\tau}=iB_t(\varepsilon+\overline{\varepsilon})r+O(r^2)=R_+(r)\label{rn4}.
\end{eqnarray}
Now, using Eq. (\ref{rn4}), we get
\begin{eqnarray}
\int\int d\varepsilon d\tau=\int\int d\varepsilon\frac{dr}{R_+(r)}=\int\int d\varepsilon\frac{dr}{\dot{r}},
\end{eqnarray}
so the action (\ref{rn3}) reads
\begin{eqnarray}
I=\int p_i
dq_i=2\int\int d\varepsilon d\tau=2\int\int d\varepsilon\frac{dr}{\dot{r}}.\label{rn8}
\end{eqnarray}
Putting Eq. (\ref{rn4}) into the above equation, and integrating around the pole at the horizon, that is, $r=0$, we get
\begin{eqnarray}
I&=&\int p_i
dq_i=2\int\int d\varepsilon\frac{dr}{\dot{r}}\nonumber\\
&=&2\int\int_{r_{in}}^{r_{out}}\frac{dr}{iB_{t}(\varepsilon+\overline{\varepsilon})r+O(r^2)}d\varepsilon\nonumber\\
&=&2\pi\int\frac{d\varepsilon}{B_{t}(\varepsilon+\overline{\varepsilon})}=-2\pi\int\frac{dM_H}{\kappa_{t}},\label{kqd}
\end{eqnarray}
where we have used the relationship of energy conservation $dM_H=-d\varepsilon$ in the last equation, and $\kappa_{t}=B_t(\varepsilon+\overline{\varepsilon})$ is the surface gravity of the WIH. So the action we considered is
\begin{eqnarray}
I=\int p_i dq_i=-2\pi\int\frac{dM_H}{\kappa_{t}}\label{rn5}.
\end{eqnarray}
As we all know
that the temperature of a black hole is proportional to the surface
gravity of the horizon
\begin{eqnarray}
T_{H}=\frac{\hbar\kappa_{t}}{2\pi}\label{rn10},
\end{eqnarray}
thus Eq. (\ref{rn5})
becomes
\begin{eqnarray}
I=\int p_i dq_i=-\hbar\int\frac{dM_H}{T_{H}}=-\hbar
\Delta S_{H}\label{rn6},
\end{eqnarray}
where we have used the first law of WIH thermodynamics (\ref{q1}) with
$\Omega_{t}=0$ in the last step, and $\Delta S_{H}<0$ is the change of entropy of WIH after emission of a particle.

At last, implementing the Bohr-Sommerfiedld quantization rule
\begin{eqnarray}
\int p_i dq_i=nh
\end{eqnarray}
in Eq. (\ref{rn6}), we derive the WIH entropy spectrum
\begin{eqnarray}
|\Delta S_{H}|=2\pi n,
\end{eqnarray}
where $n=1,2,3,...$, and it is straightforward to see that the minimum
spacing in the entropy is given by
\begin{eqnarray}
|\Delta S_{H}|=2\pi\label{rn7}.
\end{eqnarray}
Thus, the entropy spectrum is quantized and equidistant for a weakly
isolated horizon. Recalling that in the framework of Einstein's
theory of gravity, black hole entropy is proportional to the black
hole horizon area \cite{jdb}, $S_{H}=\frac{A}{4 l_{p}^2}$. It is
evident that if we employ the spacing of the entropy spectrum given
in Eq. (\ref{rn7}), the quantum of the WIH area has the form
\begin{eqnarray}
\Delta A=8\pi l_p^2\ ,
\end{eqnarray}
which is the same as the area quantum derived by Bekenstein
\cite{jdb}.

For an axial symmetric horizon, using the method in Ref.
\cite{jz}, the last equation of (\ref{rn8}) can be modified as
\begin{eqnarray}
I&=&2[\int\int d\varepsilon\frac{dr}{\dot{r}}-\int p_\phi d\phi]=2\int\int\frac{d\varepsilon-\dot{\phi}dp_{\phi}}{\dot{r}}dr\nonumber\\
&=&-2\int\int\frac{dM_H-\Omega_{t}dJ_{H}}{\dot{r}}dr=-
2\pi\int\frac{dM_H-\Omega_{t}dJ_{H}}{B_t(\varepsilon+\overline{\varepsilon})}=-\hbar
\Delta S_{H}.
\end{eqnarray}
$p_{\varphi}$ is the angular momentum of the emitted
particle, and we have used the conservation of angular momentum, that is, $dp_{\varphi}=-dJ_{H}$ and $\dot{\phi}=\Omega_{t}$ in the third equation.
In the last equality, we have used  Eq.
(\ref{rn10}) and the first law of WIH thermodynamics (\ref{q1}).  The result is the same
as that of the non-rotating WIH (\ref{rn6}).

Let us give some discussions. Firstly, Although our method is similar to the method used in Refs. \cite{bec,xxz}, there exists a significant difference. The derivation of Refs. \cite{bec,xxz} relies on the periodicity of imaginary time, while we do not need it. Secondly,  We consider a tunneling process cross the WIH, and  concentrate on the outgoing path. From the calculation, it is easy to see that the ingoing path does not contribute to the action (\ref{b1}).
Thirdly, our method considers a tunneling process and uses the
technology of the tunneling method \cite{mk1,mk3} which has been successfully used to calculate Hawking temperature of a variety of spacetimes, so our method has more generality. Finally, our method also shows that there exists close relationship between the entropy quantum and Hawking temperature, which reflects to some extent the viewpoint of the emergent perspective of gravity that temperature means existence of underlying degrees of freedom \cite{tp}.

\section{\textbf{Conclusions}}
In this paper, we have quantized the entropy and the area of a weakly isolated horizon, and obtain the quantized entropy
and area spectrum which are the same as Bekenstein' s original
results \cite{jdb}. Our results indicate that the quantization of entropy of the black hole horizon is a generic property of horizon, and is closely related to Hawking temperature.

\begin{acknowledgements}
This work is supported by National Natural Science Foundation of
China (No.11275017 and No.11173028).
\end{acknowledgements}

\end{document}